\documentclass[twocolumn,runningheads,natbib]{svjour2}
\smartqed  
\usepackage{graphicx,epsfig}
\usepackage{aas_macros}
\usepackage{txfonts}
\journalname{Astrophysics and Space Science}
\def\zerozero  {SGR 1900$+$14}
\def\zerosei  {SGR 1806$-$20}
\def\sedici {SGR 1627$-$41}
\def\zerocinque {SGR 0526$-$66}
\def \int {{\rm INTEGRAL}}
\def\flux {\mbox{erg cm$^{-2}$ s$^{-1}$}}
\def\lum {\mbox{erg s$^{-1}$}}

\def \xmm {{\rm XMM--Newton}}
\def \xte {{\rm RossiXTE}}
\def \cha {{\rm Chandra}}
\def \sax {{\rm BeppoSAX}}
\def \asca {{\rm ASCA}}
\def \batse {{\rm BATSE}}
\def \swift {{\rm Swift}}
\def \epic {{\rm EPIC}}
\def\pdot {\dot P}
\begin{document}

\title{\xmm\ observations of Soft Gamma-ray Repeaters
}


\author{
        Sandro Mereghetti \and
        Paolo Esposito  \and
        Andrea Tiengo
}

\authorrunning{S.Mereghetti, P.Esposito \& A.Tiengo} 

\institute{           S. Mereghetti, A. Tiengo \at
              INAF -- Istituto di Astrofisica Spaziale e Fisica Cosmica, Milano, Italy
          \and
          P. Esposito \at
          Universit\`a di Pavia, Dipartimento di Fisica Nucleare e Teorica and INFN-Pavia, Italy
}

\date{ Presented at the conference "Isolated Neutron Stars: 
from the Surface to the Interior", London, UK, 24-28 April 2006}

\maketitle

\begin{abstract}
All the confirmed Soft Gamma-ray Repeaters\linebreak have been
observed with the EPIC instrument on the \xmm\ satellite. We review
the results obtained in these observations, providing the most
accurate spectra on  the persistent X-ray emission in the 1--10 keV
range for these objects, and discuss them in the context of the
magnetar interpretation.
\end{abstract}

\section{Introduction}
\label{intro}

Soft Gamma-ray Repeaters (SGRs) were discovered as\linebreak sources of
short intense bursts of gamma-rays, and for a long time were
considered as a puzzling category of Gamma-ray bursts. Their
neutron star nature was immediately suggested by the 8 s
periodicity seen in the famous event of 5 March 1979, but it was
only with the discovery of their pulsating counterparts in the few
keV region that this was finally proved. Although the main
motivations for the magnetar model \citep{duncan92,thompson95} were driven by
the high energy properties of the SGRs bursts and giant flares,
X-ray observations of the ``quiescent'' emission have provided
fundamental information to understand the nature of these objects
\citep{woods04}.

Extensive observational programs have been carried out with the
\xte\ satellite, focusing mainly on the SGRs timing properties. Long
term studies based on phase-\linebreak connected timing analysis
revealed significant deviations\linebreak from a steady spin-down
\citep{woods99d,woods00,woods02}, larger than the timing noise seen
in radio pulsars and not linked in a simple way with the bursting
activity. \xte\ has also been used to investigate the variations of
the pulse profiles as a function of energy  and time
\citep{gogus02}, and to study the statistical properties of the
bursts \citep{gogus01}. However, the \xte\ observations are not
ideal to accurately measure the flux and spectrum of these
relatively faint sources located in crowded fields of the Galactic
plane, since its non-imaging instruments suffer from source
confusion and large uncertainties in the background estimate.
Imaging satellites like \sax, \asca\ and \cha\ have yielded useful
spectral information, but it is only with the advent of the large
collecting area \xmm\ satellite that high quality spectra of SGRs
have been obtained, in particular with the \epic\ instrument
\citep{struder01,turner01}.

Here we review the results obtained with the \xmm\ satellite for the
three confirmed SGRs in our Galaxy. There are also some \xmm\ data
on \zerocinque\ in the Large Magellanic Cloud, but they are of
limited use  due to the contamination from diffuse emission from the
surrounding supernova remnant and will not be discussed here.

\begin{table*}[t]
\begin{center}
\caption{\xmm\ observations and timing results for SGR 1806$-$20.
}
\label{obs}       
\begin{tabular}{cccccc}
\hline\noalign{\smallskip}
  Obs. & Date      & Duration &  Mode$^{(a)}$ and  exp. time   &
Mode$^{(a)}$ and exp. time   & Pulse Period \\
       &           &  (ks)        & PN camera   &  MOS1/2 cameras  &    (s)
\\[3pt]
 \tableheadseprule\noalign{\smallskip}
A & 2003 Apr 3   & 32  & FF (5.4 ks)  &  LW (6 ks)    & 7.5311$\pm$0.0003 \\
B & 2003 Oct 7   & 21  & FF (13.4 ks) &  LW (17 ks)    & 7.5400$\pm$0.0003\\
C & 2004 Sep 6   & 51  & SW (36.0 ks) &  LW (51 ks)    & 7.55592$\pm$0.00005 \\
D & 2004 Oct 6   & 18  & SW (12.9 ks) &  Ti (18 ks)    & 7.5570$\pm$0.0003 \\
E & 2005 Mar 7   & 24  & SW (14.7 ks) &  Ti/FF (24 ks) & 7.5604$\pm$0.0008\\
F & 2005 Oct 4   & 33  & SW (22.8 ks) &  Ti/FF (33 ks) & 7.56687$\pm$0.00003\\
G & 2006 Apr 4   & 29  & SW (20.5 ks) &  Ti/FF (29 ks) & 7.5809$\pm$0.0002\\
  \noalign{\smallskip}\hline
\end{tabular}\\
\end{center}
\begin{footnotesize}
$^{(a)}$ FF = Full Frame (time resolution 73 ms); LW = Large
Window (time resolution 0.9 s);  SW = Small Window (time
resolution 6 ms); Ti = Timing (time resolution 1.5 ms)
\end{footnotesize}
\end{table*}

\section{\zerosei}
\label{sec:zerosei}

\zerosei\ is probably the most prolific and the best studied of
the known SGRs. It showed several periods of bursting activity
since the time of its discovery in 1979\linebreak
 \citep{laros86} and
recently attracted much interest since it emitted the most powerful
giant flare ever observed from an SGR
(\citealt{hurley05,palmer05,mereghetti05}).

The low energy X-ray counterpart of \zerosei\ was identified with
the \asca\ satellite \citep{murakami94}, thanks to the detection and
precise localization of a burst simultaneously seen at higher energy
with \batse . Subsequent observations with \xte\ led to the
discovery of pulsations with P=7.5 s and $\dot
P$=8$\times$10$^{-11}$ s s$^{-1}$ \citep{kouveliotou98}.

Possible associations of \zerosei\ with the  variable non-thermal
core of a putative radio supernova remnant \citep{frail97} and with
a luminous blue variable star \citep{vankerkwijk95} were disproved
when a more precise localization of the SGR could be obtained with
the {\rm Interplanetary Network} \citep{hurleylbv99} and later
improved with \cha\ \citep{kaplan02}. The transient radio source
observed with the {\rm VLA} after the December 2004 giant flare
\citep{cameron05} led to an even smaller error region and, thanks to
the superb angular resolution (FWHM $\sim$0.1$''$) available with
adaptive optics at the {\rm ESO Very Large Telescope} a variable
near IR counterpart (K$_s$=19.3--20), could be identified
\citep{israel05}, the first one for an SGR.

The distance of \zerosei\ is subject of some debate
\citep{cameron05}, and is particularly relevant for its
implications on the total energetics of the 2004 giant flare. A
firm lower limit of 6 kpc can be derived from the HI absorption
spectrum \citep{mcclure05}, but the likely associations with a
massive molecular cloud and   with a cluster of massive stars
indicate a distance of \mbox{$\sim$15 kpc} \citep{corbel04,figer05}. In
the following we will adopt this value.

Before the \xmm\ observations, the  most accurate spectral
measurements for \zerosei\ in the soft X-ray range were obtained
with \sax\ in 1998--1999 \citep{mereghetti00}. They showed that a
power law with photon index $\Gamma$=1.95 or a thermal
bremsstrahlung with temperature kT$_{tb}$=11 keV were equally
acceptable fits. All the observations  indicated a fairly constant
flux, corresponding to a 2--10 keV luminosity of
$\sim$3$\times$10$^{35}$ \lum .

Being located at only $\sim$10$^{\circ}$ from the Galactic center
direction, \zerosei\ has been extensively observed with the \int\
satellite since 2003. A few hundreds bursts have been detected with
the {\rm IBIS} instrument in the\linebreak \mbox{15--200 keV} range,
leading to the discovery of a hardness intensity anti-correlation
and allowing to extend the number-flux relation of bursts down to
fluences smaller than\linebreak \mbox{10$^{-8}$ erg cm$^{-2}$}
\citep{gotz04,gotz06}. In addition, it was discovered with \int\
that the persistent emission from \zerosei\ extends up to 150 keV
\citep{mereghetti05int,molkov05}. The hard X-ray emission, well fit
by a power law with photon index $\Gamma\sim$1.5--1.9, seems to
correlate in hardness and intensity with the rate of burst emission,
that reached a maximum in Fall 2004.

The bursting activity of \zerosei\ culminated with the giant flare
of 2004 December 27 \citep{borkowski04,hurley05,palmer05}, that
produced the strongest flash of gamma-rays at the Earth ever
observed. The emission was so intense to cause saturation of most
in-flight detectors, significant ionization of the upper
atmosphere \citep{campbell05}, and a detectable flux of radiation
backscattered from the Moon \citep{mazets05,mereghetti05}. Other
observations of this exceptional event, and their implications for
the physics of neutron stars, are discussed elsewhere in these
proceedings \citep{stella07,israel07}. Comparing this giant flare
with those seen from \zerocinque\ and \zerozero , it is found that
the energy in the pulsating tails of the three events was roughly
of the same order ($\sim10^{44}$ ergs), while the energy in the
initial spike of \zerosei\ (a few $10^{46}$ ergs) was at least two
orders of magnitude higher than that of the other giant flares.
This indicates that the magnetic field   in the three sources is
similar. In fact the pulsating tail emission is thought to
originate from the fraction of the energy released during the
initial hard pulse that remains magnetically trapped in the
neutron star magnetosphere, forming an optically thick photon-pair
plasma \citep{thompson95}. The amount of energy that can be confined in
this way is determined by the magnetic field strength, which is
thus inferred to be of several 10$^{14}$ G in these three
magnetars.

\zerosei\ is the target of an ongoing campaign of \xmm\ observations
aimed at studying in detail the long term variations in the
properties of its persistent emission. These observations, coupled
with similar programs carried out with {\rm ESO} telescopes in the
infrared band  \citep{israel07} and at hard X-ray energy with \int\
\citep{gotz07} and {\rm Suzaku}, can be used to study the connection
between the persistent emission and the source activity level, as
manifested by the emission of bursts and flares.

\begin{figure*}
\centering
\includegraphics[width=7cm,angle=90]{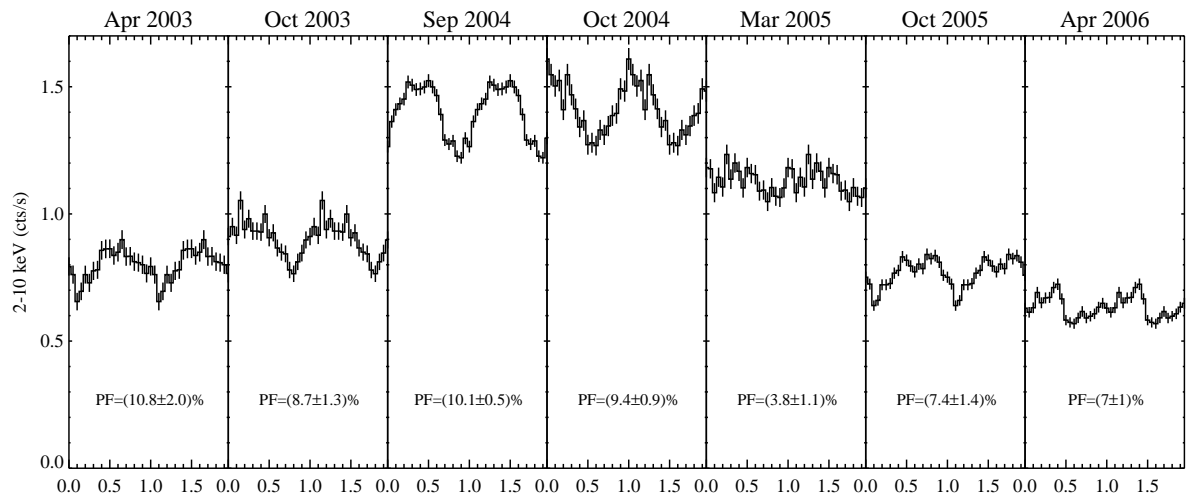}
\caption{Folded light curves of SGR 1806$-$20  obtained with the
\epic\ pn instrument in the seven \xmm\ observations. Note the flux
increase in the two observations before the December 2004 giant
flare and the small pulsed fraction in the first 2005 observation.}
\label{fig:lc}
\end{figure*}

\begin{table*}[t]
\begin{center}
\caption{Summary of the spectral results$^{(a)}$ for SGR
1806$-$20}
 \label{speres}
\begin{tabular}{ccccccc}
   \hline
Obs. & Absorption & Power Law &  kT$_{BB}$ & R$_{BB}$ &
Flux$^{(c)}$
& $\chi^2_{red}$ (d.o.f.) \\
     & 10$^{22}$ cm$^{-2}$ & photon index & (keV) & (km)$^{(b)}$
& 10$^{-11}$ erg cm$^{-2}$ s$^{-1}$  &    \\
 \hline
A & 6.6 (5.6--8.4) & 1.4 (1.0--1.7) & 0.6 (0.4--0.9) & 2.6
(0.7--13.9)
& 1.23 &  1.01 (56) \\
B & 6.0 (4.9--6.6) & 1.2 (0.5--1.4) & 0.7 (0.6--1.0) & 1.8
(1.2--2.9)
& 1.39 &  0.97 (68) \\
C & 6.5 (6.2--6.9) & 1.21 (1.09--1.35) & 0.8 (0.7--0.9) & 1.9
(1.6--2.6)
& 2.66  &  0.93 (70) \\
D & 6.5 (5.9--7.1) & 1.2 (0.9--1.4) & 0.8 (0.6--0.9) &  2.2
(1.6--3.5)
& 2.68 &  0.90 (69) \\
E & 6.0 (5.8--6.2) & 0.8 (0.5--1.0) & 0.91 (0.86--1.05) & 1.9
(1.6--2.1)
& 1.92 & 1.02 (70) \\
F & 6.4 (6.0--6.8) & 1.4 (1.1--1.7) & 0.7 (0.6--0.8) & 2.2
(1.7--3.3)
& 1.34 & 1.11 (69) \\
G & 6.2 (5.6--6.6) & 1.2 (0.9--1.4) & 0.7 (0.6--0.8) & 2.0
(1.6--2.7)
&  1.07 & 1.09 (68) \\
\hline
\end{tabular}
\end{center}
\begin{small}
$^{(a)}$ Errors are at the  90\% c.l. for a single interesting parameter \\
$^{(b)}$ Radius at infinity assuming a distance of 15 kpc \\
$^{(c)}$ Absorbed flux in the 2-10 keV energy range \\
\end{small}
\end{table*}

\subsection{\xmm\ results}

Seven \xmm\ observations of \zerosei\ have been carried out to
date (see Table~\ref{obs}). Four were obtained from April 2003 to
October 2004, before the giant flare\linebreak \citep{mte05}. At
the time of the giant flare the source was not visible by \xmm\
(and most other satellites) due to its proximity to the Sun, thus
the next observation could be done only in March 2005
\citep{tiengo05}. This was followed by another observation in
October \citep{rea05} and a most recent one in April 2006, the
results of which are presented here for the first time.

The bursts detected in some of these observations (mostly in
September-October 2004) were excluded by appropriate time selections
to derive the spectral results reported below. After screening out
the bursts, the source pulsations were clearly detected in all the
observations. The corresponding folded light curves are shown in
Fig.~\ref{fig:lc}, where all the panels have the same scale in count
rate to facilitate a comparison of the flux variations between the
observations. The main spectral results are summarized in
Table~\ref{speres}, where we have reported only the best fit
parameters for the power law plus blackbody model (see
\citealt{mte05,tiengo05} for more details).

Indeed the strong requirement for a blackbody  component is one of
the main results of the high quality \xmm\ spectra. In this
respect the most compelling evidence comes  from the September
2004 observation (obs. C), which,  thanks to the high source count
rate and long observing time, provided the spectra with the best
statistics.  A fit with an absorbed power law yields a relatively
high $\chi^{2}$ value ($\chi^{2}_{red}$=1.37) and structured
residuals, while a much better fit ($\chi^{2}_{red}$=0.93) can be
obtained by adding a blackbody component. The best fit parameters
are photon index $\Gamma$=1.2, blackbody temperature kT$_{BB}$=0.8
keV and absorption N$_H\sim$6.5$\times$10$^{22}$ cm$^{-2}$.
Although some of the observations with lower statistics give
acceptable fits also with a single power law, the results reported
in Table~\ref{speres} indicate that all the observations are
consistent with the presence of an additional blackbody component
with similar parameters.

The second new result derived from these observations is the long
term flux variability. All the observations   of \zerosei\ in the
1--10 keV range obtained in the previous years with {\rm ROSAT},
\asca\ and \sax\ were consistent with a flux of $\sim$$10^{-11}$ erg
cm$^{-2}$ s$^{-1}$.  On the other hand, the \xmm\ data showed a
doubling of the flux in September-October 2004 followed by a gradual
recovery to the ``historical'' level during the observations
performed after the giant flare. Interestingly, the same trend was
seen above 20 keV with \int\ \citep{mereghetti05int,gotz07}, as well
as in the flux of the NIR counterpart \citep{israel05,israel07}.

\begin{figure}
\centering
\includegraphics[width=5.5cm,angle=-90]{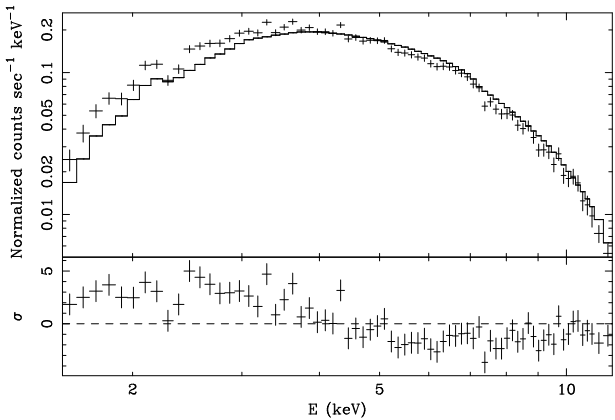}
\caption{Top panel: the data represent the spectrum of obs. G,
while the line is the best fit model of obs. C simply rescaled in
normalization. The residuals shown in the bottom panel clearly
indicate that the spectrum was softer after the giant flare.}
\label{spec}
\end{figure}

The observations performed before and after the giant flare show
significant differences also in the source pulsed fraction and
spectral shape. The pulsed fraction in the first observation after
the flare was the smallest seen with \xmm , while it increased
again in the following observations. The spectral hardness
followed a similar trend: the four pre-flare observations give
marginal evidence for a gradual hardening, while the spectrum was
definitely softer in the post-flare observations. This is
illustrated in Fig.~\ref{spec}, which shows the April 2006
spectrum fitted with the pre-flare model (obs. C): the trend in
the residuals clearly indicate the spectral softening.

Finally, we can compare the \xmm\ spectra with those obtained in the
previous years with other satellites. We base this comparison on the
fits with a single power law, since this model  was successfully
used to describe the \asca\ and \sax\ data. The average photon index
in the four \xmm\ observations of 2003--2004 ($\Gamma$=1.5 $\pm$0.1)
was significantly smaller than that observed in 1993 with \asca\
($\Gamma$=2.25$\pm$0.15, \citealt{mereghettisax02}) and in
1998--2001 with \sax\ ($\Gamma$=1.97$\pm$0.09,
\citealt{mereghettisax02}). This indicates that a  spectral
hardening occurred between September 2001 and April 2003.

A long term variation occurred also in the average spin-down rate:
while the early sparse period measurements with \asca\ and \sax\
\citep{mereghettisax02}, as well as a phase-connected \xte\ timing
solution spanning\linebreak February-August 1999 \citep{woods00},
were consistent with an average $\pdot\sim8.5\times10^{-11}$ s
s$^{-1}$, subsequent  \xte\  data indicate a spin-down larger by a
factor $\sim$4 \citep{woods02} and the four \xmm\ period
measurements before the giant flare  show a further increase to an
average $\pdot$=5.5$\times$10$^{-10}$ s s$^{-1}$ \citep{mte05}.

\begin{figure}
\centering
\includegraphics[width=8cm,angle=00]{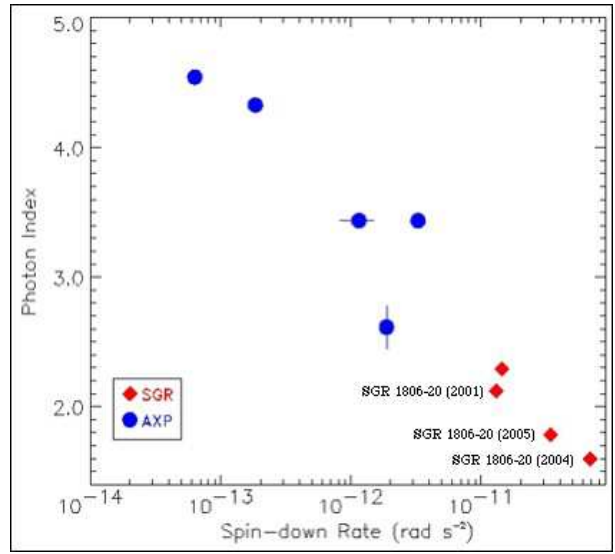}
\caption{Correlation between power law photon index and spin-down
rate in Anomalous X-ray Pulsars and SGRs (adapted from
\citealt{marsden01}). Each point refers to a different source,
except the three points for SGR 1806$-$20 in different time
periods.} \label{corr}
\end{figure}

As shown in Fig.~\ref{corr}, the changes in spectral hardness and
spin-down rate of \zerosei\ follow the correlation between these
quantities discovered in the sample of AXPs and SGRs by comparing
different objects: the sources with the harder spectrum have a
larger long term spin-down rate \citep{marsden01}. These results
indicate, for the first time,  that such a correlation also holds
within different states of a single source.

\section{\zerozero}

Bursts from this SGR were  discovered  with the {\rm Venera}
satellites in 1979  \citep{mazets79}. No other bursts were detected
until thirteen years later, when four more events were seen with
\batse\ in 1992 \citep{kouveliotou93}. In the meantime the X-ray
counterpart had been discovered with {\rm ROSAT} \citep{vasisht94},
and later found to pulsate at 5.2 s with \asca\ \citep{hurley99b}.
Subsequent observations with the \xte\ satellite confirmed the
pulsations and established that the source was spinning down
rapidly, with a period derivative of $\sim$$10^{-11}$ \mbox{s
s$^{-1}$} \citep{kouveliotou99}.

The peak of the activity  from \zerozero\ was reached on 1998
August 27, with the emission of  a giant flare
\citep{hurley99gf,feroci99}, resembling the only similar event
known at that time, the exceptional burst of 5 March 1979. The
1998 giant flare from \zerozero\ could be studied much better than
that of \zerocinque . The flare started with a short ($\sim$0.07
s) soft spike, followed by a much brighter short and hard pulse
that reached a peak luminosity of $\sim$$10^{45}$ \mbox{erg
s$^{-1}$}. The initial spike was followed by a softer gamma-ray
tail modulated at 5.2 s \citep{hurley99gf,mazets99}, which decayed
in a quasi exponential manner over the next $\sim$6 minutes
\citep{feroci01}. Integrating over the entire flare assuming
isotropic emission, at least $10^{44}$ erg were released in hard
X-rays above 15 keV \citep{mazets99}. \zerozero\ also emitted
another less intense flare on 18 April 2001
\citep{feroci03,guidorzi04}, which based on its energetics was
classified as an ``intermediate'' flare.

Despite being the less absorbed of the galactic SGRs (\mbox{N$_H$
$\sim2\times10^{22}$ cm$^{-2}$}) no optical/IR counterpart has been
yet identified for \zerozero . Its possible association with a
young cluster of massive stars \citep{vrba00}, where the SGR could
have been born, gives a distance of $\sim$15 kpc, that we will
adopt in the following.

Recently, persistent emission from \zerozero\ has been detected also
in the 20--100 keV range thanks to observations with the \int\
satellite \citep{gotz06}.

\subsection{\xmm\ results}

\zerozero\ lies in a sky region that, until recently, was not
observable by \xmm\ due to technical constraints in the satellite
pointing. Thus the first observation of \zerozero\ could be obtained
only in September 2005\linebreak \citep{met06}. This observation
occurred during a long period of inactivity (the last bursts before
the observations were reported in November 2002,
\citealt{hurley02}).

The spectrum could not be fit satisfactorily with single component
models, while a good fit was obtained with the sum of a power law
and a blackbody, with  photon index $\Gamma$=1.9$\pm$0.1,
temperature kT=0.47$\pm$0.02 keV,  absorption\linebreak \mbox{N$_H$
= $(2.12\pm0.08)\times10^{22}$ cm$^{-2}$}, and unabsorbed
flux\linebreak \mbox{$\sim$$4.8\times10^{-12}$ \flux} (2--10 keV).
An acceptable fit\linebreak could also be obtained with the sum of
two blackbodies with temperatures of 0.53 and 1.9 keV.

The \xmm\ power law plus blackbody parameters are in agreement with
previous observations of this source carried out with \asca\
\citep{hurley99b}, \sax\  \citep{woods99b,esposito06} and \cha\
\citep{kouveliotou01}, but the flux  measured in\linebreak September
2005 is the lowest ever seen from \zerozero. A $\sim$30\% decrease
of the persistent emission, compared to the ``historical'' level of
$\sim$$10^{-11}$ \flux, had already been noticed in the last \sax\
observation \citep{esposito06}, that was carried out in April 2002,
six month earlier than the last  bursts reported before the recent
reactivation. The long term fading experienced by \zerozero\ in
2002--2005 might be related to the apparent decrease in the bursting
activity in this period.

A second \xmm\ observation was carried out on 1 April 2006, as a
target of opportunity following the source reactivation indicated by
a few bursts detected  by \swift\ \citep{palmer06} and {\rm
Konus-Wind} \citep{golenetskii06}. The spectral shape was consistent
with that measured in the first observation, but the flux was
$\sim$$5.5\times10^{-12}$ \flux~\citep{met06}.

The spin period was  \mbox{$5.198346\pm0.000003$ s} in September
2005 and \mbox{$5.19987\pm0.00007$ s} in April 2006. In both
observations the pulsed fraction was $\sim$16\% and no significant
changes in the pulse profile shape were seen after the burst
reactivation.

For both observations we performed  phase-resolved spectroscopy
extracting the  spectra for different selections of\linebreak phase
intervals. No significant variations with phase were detected, all
the spectra being consistent with the model and parameters of the
phase-averaged spectrum, simply rescaled  in normalization.

\begin{table*}[t]
\caption{Main properties of the four confirmed SGRs}
\label{tab:1}       
\begin{tabular}{|l|c|c|c|c|}
\hline\noalign{\smallskip}
     & \sedici\       & \zerosei\  &  \zerozero\ & \zerocinque\  \\
 \tableheadseprule\noalign{\smallskip}
Coordinates &  16$^h$ 35$^m$ 51.83$^s$  & 18$^h$ 08$^m$ 39.337$^s$  & 19$^h$ 07$^m$ 14.33$^s$  &  05$^h$ 26$^m$ 00.89$^s$    \\
            &  $-$47$^\circ$ 35$'$ 23.3$''$  & $-$20$^\circ$ 24$'$ 39.85$''$  & $+$09$^\circ$ 1$'$9 20.1$''$  & $-$66$^\circ$ 04$'$ 36.3$''$     \\
Error       &  0.2$''$ $[a]$  & 0.06$''$ $[b]$  & 0.15$''$ $[d]$  &  0.6$''$ $[g]$     \\
                   & & & & \\
Distance   &  11 kpc  & 15 kpc  &  15 kpc&  50 kpc     \\
Period    &  --  & 7.6 s & 5.2 s &  8 s    \\
Period derivative (s s$^{-1}$) & --   & $(8.3-81)\times10^{-11}$ $[c]$  & $(6.1-20)\times10^{-11}$ $[e]$  &   $6.6\times10^{-11}$ $[g]$    \\
Magnetic field$^a$ & --  &  $(8-25)\times10^{14}$ G & $(6-10)\times10^{14}$ G &  $7\times10^{14}$ G    \\
&    &   &   &       \\
Flux range$^b$ (erg cm$^{-2}$ s$^{-1}$) & $(0.025-0.6)\times10^{-11}$   & $(1.3-3.8)\times10^{-11}$  & $(0.5-2.7)\times10^{-11}$  &  $0.07\times10^{-11}$     \\
Typical flux$^b$ (erg cm$^{-2}$ s$^{-1}$) & $\sim3\times10^{-13}$   & $\sim1.5\times10^{-11}$  & $\sim10^{-11}$  &  $\sim10^{-12}$     \\
20--60 keV flux (erg cm$^{-2}$ s$^{-1}$) & --   & $(3-5)\times10^{-11}$  & $\sim1.5\times10^{-11}$  &  --     \\
  &   &  &   &  \\
Optical/IR & J$>$21.5, H$>$19.5,  & K$_s$=20--19.3 $[b]$  &  J$>$22.8, K$_s>$20.8 $[f]$ &  V$>$27.1, I$>$25 $[h]$   \\
           & K$_s>$20.0 $[a]$  & J$>$21.2, H$>$19.5  &  &       \\
Luminosity$^c$ (erg s$^{-1}$) & $\sim4\times10^{33}$   & $\sim4\times10^{35}$  &  $\sim3\times10^{35}$ &    $\sim2\times10^{35}$   \\
Photon index  & 3  & 1.2  & 2  & 3.1 \\
Blackbody kT   & --  & 0.8  keV &  0.45  keV&  0.53  keV\\
N$_H$         & $9\times10^{22}$ cm$^{-2}$  & $6.5\times10^{22}$ cm$^{-2}$  &  $2.2\times10^{22}$ cm$^{-2}$ & $0.55\times10^{22}$ cm$^{-2}$  \\
  &   &  &   &  \\
   &    &   &   &  \\
Giant Flare &  --  & December 27, 2004  &  August 27, 1998 &   March 5, 1979    \\
Initial spike energy (erg) & --  & $(1.6-5)\times10^{46}$  & $>6.8\times10^{43}$  & $1.6\times10^{44}$      \\
Pulsating tail energy (erg) &  --  &  $1.3\times10^{44}$  & $5.2\times10^{43}$   &   $3.6\times10^{44}$    \\
Most active periods & 1998 Jun--Jul & 1983--1985, 1996--1999, & 1979 Mar, 1992, 1998--1999, & 1979 Mar--Apr, \\
   & &  2003--2004  & 2001--2002, 2006   & 1981 Dec--1983 Apr \\

   \noalign{\smallskip}\hline
\end{tabular}
\textsc{References:}\\
\begin{tabular}{llll}
$[a]$ \citep{wachter04} & $[c]$ \citep{woods06} & $[e]$ \citep{woods02}& $[g]$ \citep{kulkarni03}\\
$[b]$ \citep{israel05}  & $[d]$ \citep{frail99} & $[f]$ \citep{kaplan02}& $[h]$ \citep{kaplan01}\\
\end{tabular}\\
\textsc{Notes:}\\
$^{(a)}$ Assuming spin-down due to dipole radiation:
$B=3.2\times10^{19}(P\dot{P})^{1/2}~\rm{G}$\\
$^{(b)}$ Unabsorbed flux in the 2--10 keV energy range\\
$^{(c)}$ Luminosity in the 2--10 keV energy range assuming the
distances reported above\\
\end{table*}

\section{\sedici}
\label{sec:sedici}

From the point of view of the bursts and timing properties,
\sedici\ is one of the less well studied SGRs. This source was
discovered during a period of bursting activity that lasted only
six weeks in 1998 \citep{woods99c,hurley99,mazets99}. Since then no
other bursts were observed.

With a column density of N$_H\sim10^{23}$ cm$^{-2}$, corresponding
to A$_V$$\sim$40--50, \sedici\ is the most absorbed of the known
SGRs. Thus it is not surprising that little is known on its
possible counterparts. Near IR observations \citep{wachter04}
revealed a few objects positionally consistent with the small
\cha\ error region, but they are likely foreground objects
unrelated to the SGR.

A distance of 11 kpc  is generally  assumed for\linebreak \sedici,  based on
its possible association with the radio complex CTB 33, comprising
the supernova remnant SNR G337.0$-$0.1 and a few HII regions
\citep{corbel99}.

During the active period a soft X-ray counterpart with  flux
F$_x\sim7\times10^{-12}$ \flux\ was identified with \sax\
\citep{woods99c}. However it was not possible to reliably measure a
periodicity (a marginal detection at 6.4 s was not confirmed by
better data).

Observations carried out in the following years with \sax, \asca\
and \cha\, showed a monotonic decrease in the luminosity, from the
value of $\sim$10$^{35}$ \lum\ (for d=11 kpc) seen in 1998 down to
$\sim$4$\times$10$^{33}$ erg s$^{-1}$.

\begin{figure}
\centering
\includegraphics[width=8cm,angle=90]{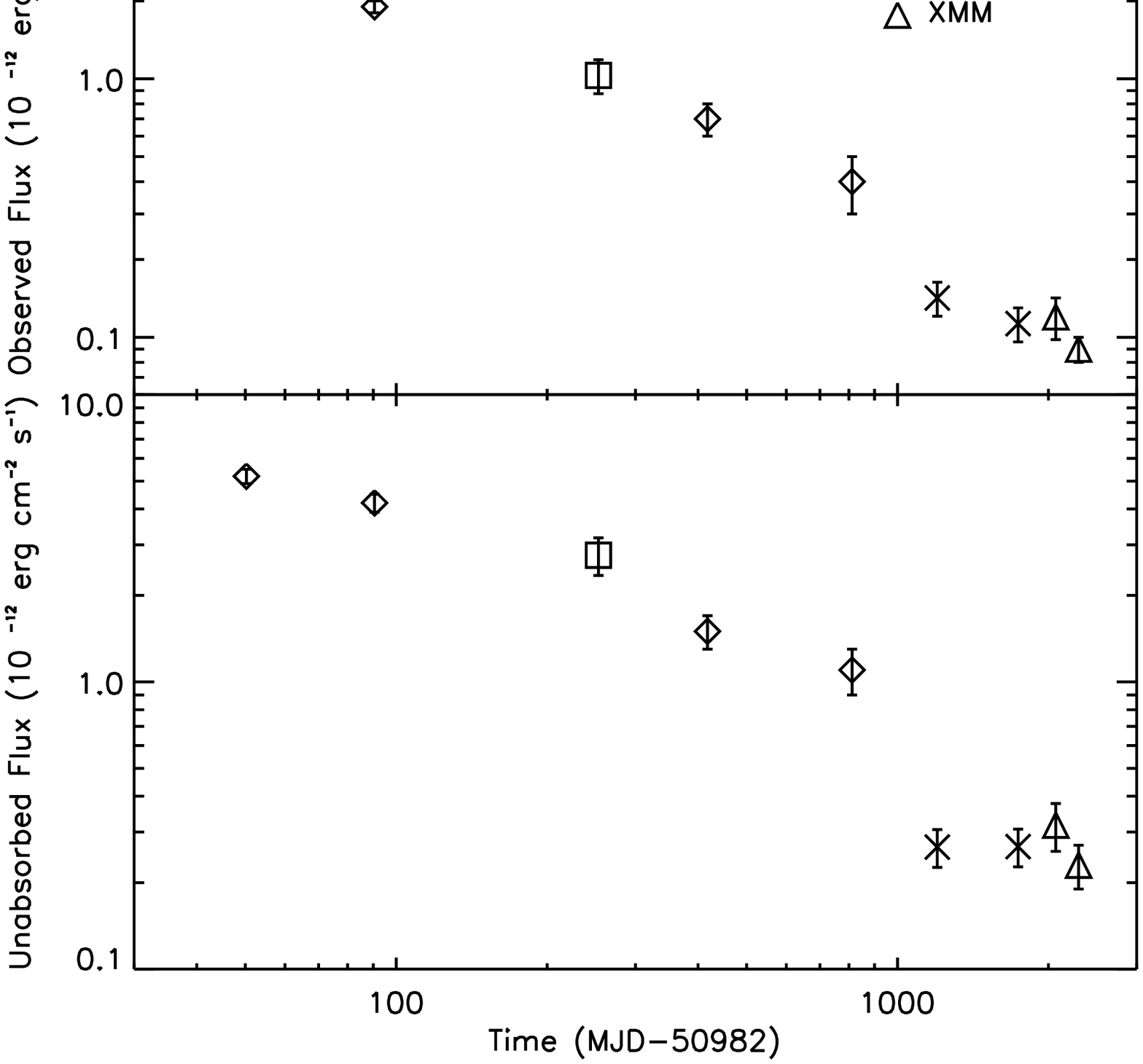}
\caption{ Long term light curve of \sedici\  based on data from
different satellites \citep{mereghetti06}.  Top panel: absorbed
flux in the 2--10 keV range. Bottom panel: unabsorbed flux in the
2--10 keV range. For clarity, the \xmm\ points of 2004 September 4,
which are consistent with the last measurement, are not plotted. }
 \label{fig:1627}
\end{figure}

\subsection{\xmm\ results}

A 52 ks long \xmm\ pointing on  \sedici\ was done in September
2004 \citep{mereghetti06}, while some other information could be extracted from two
observations in which \sedici\ was serendipitously detected at an
off-axis angle of $\sim10'$. These $\sim$30 ks long observations,
whose main target was IGR J16358--4726, were carried out in
February and September 2004.

All the \xmm\ data showed a rather faint source
($\sim$9$\times10^{-14}$ erg cm$^{-2}$ s$^{-1}$) with a soft
spectrum. Unfortunately the source faintness did not allow a
sensitive search for periodic pulsations. Both a steep power law
(photon index\linebreak \mbox{$\Gamma=3.7\pm0.5$}) and a blackbody with temperature\linebreak
\mbox{kT$_{BB}=0.8^{+0.2}_{-0.1}$ keV} gave acceptable fits. The
absorption was consistent with that measured in all the previous
observations, N$_H$=9$\times10^{22}$ cm$^{-2}$. There is evidence
that the spectrum softened between the two \cha\ observations
carried out in September 2001 and August 2002 \citep{kouveliotou03}. The
photon index measured with \xmm\ is consistent with that of the
last \cha\ observation but, due to the large uncertainties, also a
further softening cannot be excluded.

The \xmm\ flux measurements are compared with those obtained in
previous observations in Fig.~\ref{fig:1627}. The two panels refer
to the observed (top) and emitted (bottom) fluxes in the 2--10 keV
range and for a common value of the absorption in all the
observations (N$_H$=9$\times10^{22}$ cm$^{-2}$, see\linebreak
\citealt{mereghetti06} for details). The long term decrease in
luminosity is clear, but, owing to the source spectral variations,
the details of the decay light curve are different for the observed
and unabsorbed flux.

If one considers the observed fluxes, the \cha\ and \xmm\ data
suggest that \sedici\ continued to fade also after September 2001,
while the unabsorbed values indicate a possible plateau level at
$\sim$2.5 $\times10^{-13}$ \flux . It is important to realize that
the quantity most relevant for theoretical modeling, i.e. the
emitted flux, is subject to the uncertainties in the spectral
parameters. This is particularly important for high N$_H$ values
and small fluxes, as in the case of \sedici .

The long term luminosity decrease of \sedici\ was interpreted as
evidence for cooling of the neutron star surface after the deep
crustal heating that occurred during the period of SGR activity in 1998. The
decay light curve was fitted with a model of deep crustal heating
requiring a massive neutron star (M$>$1.5 $M{_\odot}$) which could
well explain a plateau seen between days 400 and 800
\citep{kouveliotou03}. However, the evidence for such a plateau is
not so compelling, according to our reanalysis of the \sax\ data. In
fact all the \sax\ and \asca\ points in the top panel of
Fig.~\ref{fig:1627}, before the rapid decline seen with \cha\ in
September 2001, are well fit by a power law decay,
F(t)$\propto$(t$-$t$_0$)$^{-0.6}$, where t$_0$ is the time of the
discovery outburst.

\section{\xmm\ results on the SGRs bursts}

Up to now limited spectral information has been obtained for SGR
bursts below 20 keV. In particular, before our \xmm\ observations,
spectra with good energy resolution and sensitivity at a few keV
were lacking. However, some studies have provided evidence that the
optically thin thermal bremsstrahlung model, which gives a good
phenomenological description of the burst spectra in the hard X-ray
range, is inconsistent with the data below \mbox{15 keV}
\citep{fenimore94,olive04,feroci04}. We therefore tried to address
this issue using the \xmm\ data.

Several tens of bursts were detected during some of the \xmm\
observations of \zerosei\ (while none was seen in \zerozero\ and
\sedici).  These bursts had durations typical of the short bursts
more commonly observed at higher energy. Since the individual bursts
had too few counts for a meaningful spectral analysis, we extracted
a cumulative spectrum by summing all the bursts detected during the
2004 observations. The resulting spectrum corresponds to a total
exposure of 12.7 s and contains about 2000 net counts in the 2--10
keV range. We checked that pile-up effects were not important (see
\citealt{mte05} for details). The spectrum of the remaining
observing time was used as background.

All the fits with simple models (power law, blackbody, thermal
bremsstrahlung) gave formally acceptable $\chi^{2}$ values, but the
power law and the bremsstrahlung required a large absorption
(N$_H$=10$^{23}$ cm$^{-2}$), inconsistent with the value seen in the
persistent emission.  We therefore favor the blackbody model, which
yields kT$_{BB}$=2.3$\pm$0.2 keV and N$_H$=6$\times$10$^{22}$
cm$^{-2}$, in agreement with the value determined from the spectrum
of the persistent emission.

The residuals from this best fit showed a deviation at \mbox{4.2 keV}.
Although the deviation is formally at 3.3$\sigma$, it could not be
reproduced in spectra obtained with different data selections and
binning criteria. Therefore we consider it as only a marginal
evidence for an absorption line \citep{mte05}.

\begin{figure}
\centering
\includegraphics[width=8cm,angle=00]{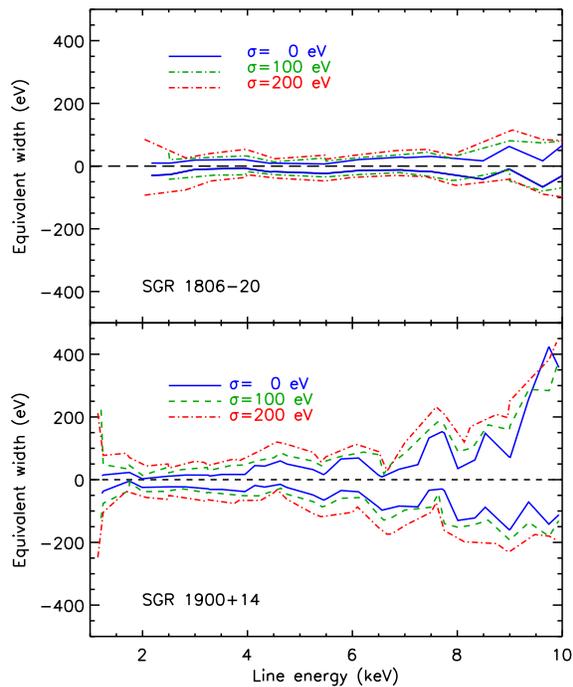}
\caption{Upper limits (at 3$\sigma$) on spectral features in the
persistent emission of \zerosei\ (top) and \zerozero\ (bottom). }
\label{fig:lines}
\end{figure}

\section{(Absence of) spectral lines}
In models involving ultra-magnetized neutron stars, proton cyclotron
features are expected to lie in the X-ray range, for surface
magnetic fields strengths of $\sim 10^{14}-10^{15}$~G. Detailed
calculations of the spectrum emerging from the atmospheres of
magnetars in quiescence have confirmed this basic expectation
\citep{zane01,ho01}. Model spectra exhibit a strong absorption line
at the proton cyclotron resonance, $E_{c,p}\simeq 0.63
z_G(B/10^{14}\, {\rm G})$ keV, where $z_G$, typically in the
0.70--0.85 range, is the gravitational red-shift at the neutron star
surface. No evidence for persistent cyclotron features has been
reported to date in SGRs, despite some features have been possibly
detected during bursts (see e.g. \citealt{strohmayer00,ibrahim03}).

A sensitive search for spectral lines was among the main
objectives of the \xmm\ observations of SGRs. However no evidence
for emission or  absorption lines, was found by looking at the
residuals from the best fit models.  In the case of \zerosei\ and
\zerozero\ the upper limits are the most constraining ever
obtained for these sources in the $\sim$1--10 keV energy range.
They are shown in Fig.~\ref{fig:lines} for the most sensitive
observation of each source, i.e. those of September 2004 for
\zerosei\ and of September 2005  for \zerozero . The plotted
curves represent the upper limits on the equivalent widths as a
function of the assumed line energy and width. They were derived
by adding Gaussian components to the best fit models and computing
the allowed range in their normalization.

\begin{figure}
\centering
\includegraphics[width=9cm,angle=00]{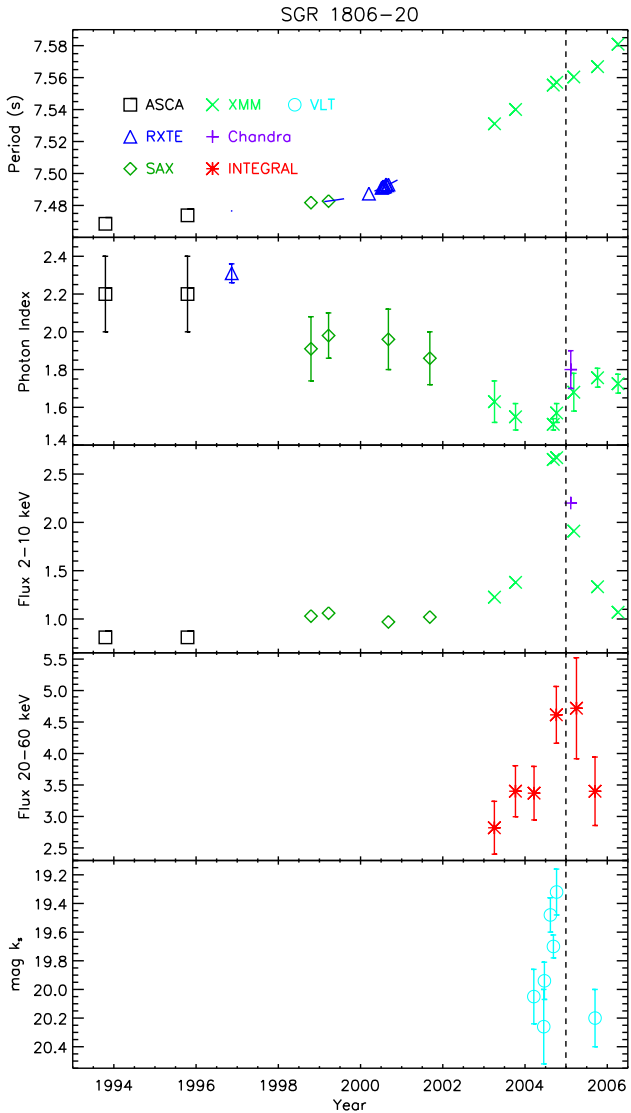}
\caption{From top to bottom: long term evolution of the pulse
period, photon index, X-ray flux (2--10 keV), hard X-ray flux
(20--60 keV), and infrared magnitude of \zerosei . Fluxes are in
units of 10$^{-11}$ erg cm$^{-2}$ s$^{-1}$. The vertical dashed
line indicates the December 2004 giant flare. } \label{1806}
\end{figure}

\begin{figure}
\centering
\includegraphics[width=9cm,angle=00]{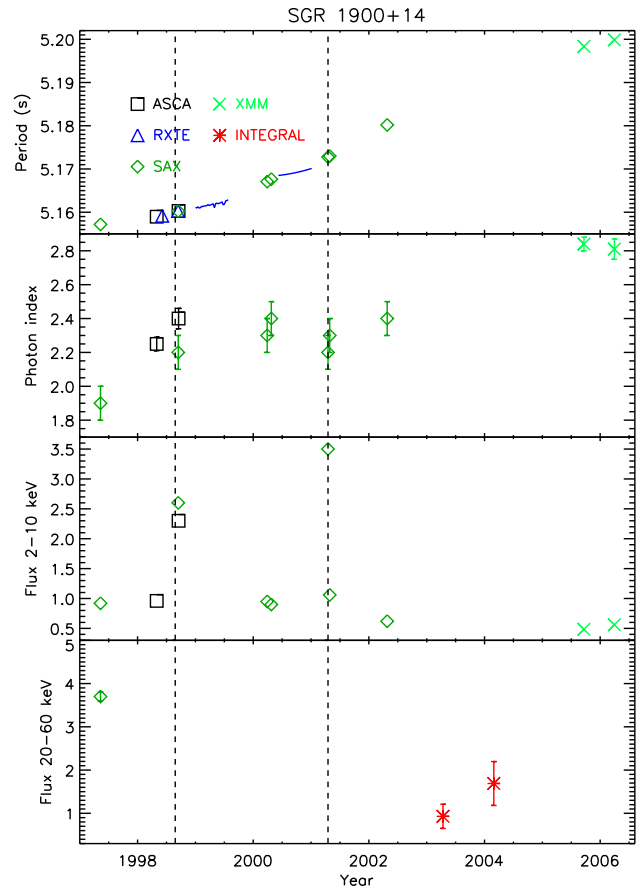}
\caption{From top to bottom: long term evolution of the pulse
period, photon index, X-ray flux (2--10 keV), and hard X-ray flux
(20--60 keV) of \zerozero . Fluxes are in units of 10$^{-11}$ erg
cm$^{-2}$ s$^{-1}$.   The vertical dashed lines indicate the 27
August 1998 giant flare and the 18 April 2001 intermediate flare.
} \label{1900}
\end{figure}

Some reasons have been proposed to explain the absence of
cyclotron lines in magnetars, besides the obvious possibility that
they lie outside the sampled energy range. Magnetars might differ
from ordinary radio pulsars because their magnetospheres are
highly twisted and therefore can support current flows
\citep{tlk02}. The presence of charged particles ($e^-$ and ions)
produces a large resonant scattering depth and since  i) the
electron distribution is spatially extended and ii) the resonant
frequency depends on the local value of the magnetic field,
repeated scatterings could lead to the formation of a hard tail
instead of a narrow line.  A different explanation involves vacuum
polarization effects. It has been  calculated that in strongly
magnetized atmospheres this effect can significantly reduce the
equivalent width of cyclotron lines, thus making difficult their
detection \citep{ho03}.


\section{Conclusions}

Many of the results presented above fit reasonably well with the
magnetar model interpretation. However, there are also a few
aspects that require more theoretical and observational efforts to
be interpreted in this framework, in particular when one considers
the variety of different behaviors shown by these sources and
their close relatives like the Anomalous X-ray pulsars
\citep{kaspi07}.

The long term variations seen in \zerosei , the\linebreak source observed
more often with \xmm , are qualitatively consistent with the
predictions of the magnetar model involving a twisted
magnetosphere \citep{tlk02}.
As mentioned above, according to this model, resonant scattering
from magnetospheric currents  leads to the formation of a
high-energy tail. A gradually increasing twist results in a larger
optical depth that causes a hardening of the X-ray spectrum.  At the
same time, the spin down rate increases because, for a fixed dipole
field, the fraction of field lines that open out across the speed of
light cylinder grows.  In addition, the stresses building up in the
neutron star crust and the magnetic footprints movements lead to
crustal fractures causing an increase in the bursting activity.
Since spectral hardening, spin-down rate, and bursting rate increase
with the twist angle, it is not surprising that these quantities varied in
a correlated way in \zerosei\ (see Fig.~\ref{1806} and
\citealt{gotz07}). However, as visible in Fig.~\ref{1806}, while the
spectral hardening took place gradually over several years, the
spin-down variation occurred more rapidly in 2000. A recent analysis
of \xte\ data around the time of the giant flare \citep{woods06}
shows that the correlation between spectral and variability
parameters is indeed rather complex.

The long term flux evolution of  \zerozero\ is shown in
Fig.~\ref{1900}. It can be seen that, excluding the enhancements
seen in correspondence of the flares, the luminosity remained
always at the same level in the years 1997--2001, and then decreased
slightly until the lowest value seen with \xmm\ in September 2005.
The following observation of April 2006 showed that the decreasing
luminosity trend has been interrupted by the recent onset of
bursts emission. However, the moderate flux increase was not
associated with significant changes in the X-ray spectral and
timing properties, probably because the source is, up to now, only
moderately active.

The  luminosity now reached by \sedici ,\linebreak
\mbox{$\sim$3.5$\times$10$^{33}$ erg s$^{-1}$} is the smallest ever
observed from a SGR, and is similar to that of the low state of the
transient anomalous X-ray pulsar XTE J1810$-$197
\citep{ibrahim04,gotthelf04}. This low luminosity might be related
to the long period ($\sim$6 years) during which \sedici\ has not
emitted bursts. However, the behavior of this source differs from
that of the other SGRs that during periods of apparent lack of
bursts changed only moderately their luminosity. In fact no bursts
were detected from \zerozero\ in the three years preceding the
\xmm\ observations: had its luminosity decreased with the same
trend exhibited by \sedici\ it would have been much fainter than
observed by \xmm . Even more striking is the case of \zerocinque ,
which has a high luminosity  ($\sim2\times10^{35}$ erg s$^{-1}$),
despite no signs of  bursting activity have been observed in the
last 15 years  (unless weak bursts from this source have passed
undetected due to its larger distance and location in a poorly
monitored sky region).

It thus seems that, similarly to the case of Anomalous X-ray
Pulsars, SGRs comprise both  ``persistent'' sources (\zerosei ,
\zerozero\ and \zerocinque\ ) and ``transients'' (\sedici\ ). The
reasons behind this difference are currently unclear and not
necessarily the same that differentiate between SGRs and AXPs.

\begin{acknowledgements}
This work has been supported by the Italian Space Agency through
the contract ASI-INAF I/023/05/0. XMM-Newton is
an ESA science mission with instruments and contributions
directly funded by ESA Member States and NASA.
\end{acknowledgements}

\bibliographystyle{aa}


\end{document}